\documentclass[traditabstract]{aa} 

\usepackage{graphicx}
\usepackage{longtable,lscape}
\usepackage{natbib}
\bibpunct{(}{)}{;}{a}{}{,}
\usepackage[usenames]{color}

\usepackage{txfonts}
\begin{document}

\title{On the white dwarf cooling sequence with extremely large telescopes}

\author{G. Bono\inst{1,2}\fnmsep\thanks{\email{bono@roma2.infn.it}}
\and
M. Salaris\inst{3}
\and
R. Gilmozzi\inst{4}	  
}

\offprints{G. Bono}

\institute{
Dipartimento di Fisica, Universita' di Roma Tor Vergata Via Della Ricerca Scientifica 1, 00133 Roma, Italy
\and
INAF, Rome Astronomical Observatory, Via Frascati 33, 00040 Monte Porzio Catone, Italy
\and
Astrophysics Research Institute, Liverpool John Moores University, 12 Quays House, Birkenhead CH41 1LD
\and
European Southern Observatory, Karl-Schwarzschild-Str. 2, 85748 Garching bei Munchen, Germany
}

\date{Received  ; accepted }

\abstract{We present new diagnostics of white dwarf (WD) cooling sequences and luminosity
functions (LFs) in the near-infrared (NIR) bands that will exploit the sensitivity
and resolution of future extremely large telescopes. The collision-induced absorption
(CIA) of molecular hydrogen causes a clearly defined blue turn-off along the WD (WDBTO) 
cooling sequences and a bright secondary maximum in the WD LFs. These features are independent 
of age over a broad age range and are minimally affected by metal abundance. This means
that the NIR magnitudes of the WDBTO are very promising distance indicators. The interplay
between the cooling time of progressively more massive WDs and the onset of CIA causes
a red turn-off along the WD (WDRTO) cooling sequences and a well defined faint peak in the
WD LFs. These features are very sensitive to the cluster age, and indeed the K-band
magnitude of the faint peak increases by 0.2--0.25 mag/Gyr for ages between 10 and 14 Gyr.
On the other hand, the faint peak in the optical WD LF increases in the same age range
by 0.17 (V band) and 0.15 (I band) mag/Gyr.
Moreover, we also suggest to use the difference in magnitude between the main sequence
turn-off and either the WDBTO (optical) or the WDRTO (NIR). This age diagnostic
is also independent of distance and reddening. Once again the sensitivity in the K band
(0.15-0.20 mag/Gyr) is on average a factor of two higher than in the optical bands
(0.10 [V band], 0.07 [I band] mag/Gyr).
Finally, we also outline the use of the new diagnostics to constrain the CO phase
separation upon crystallization.} 


\maketitle

%

\section{Introduction}\label{introduction}

Among the several relevant contributions by A. Sandage to stellar astrophysics,
the absolute age of Galactic globular clusters (GGCs) has played a major role in
astrophysical research. Back in the early fifties he compared the main-sequence 
turn-off (MSTO) stars in optical color-magnitude diagrams (CMDs)
with early cluster isochrones to constrain the age difference both between
different clusters \citep{san53,san54} and with the formation and evolution
of the Galactic Spheroid \citep{egg62}.

This approach has been and still is the most popular method to constrain
the age of Galactic \citep{buo98,ste99,sal02,gra03,zoc03,dea05,dic10}
and extragalactic \citep{bro04,mil09} stellar systems.
Related uncertainties involve the distance modulus (derived either
empirically or from theoretical horizontal branch models), the reddening
(depending on the methodology followed \citep{ren91,bon08}),
the GGC total metallicity \citep{kra03,car09,asp09}, the photometric
%
%
precision \citep{ste05}, and the efficiency of atomic diffusion in
stellar interiors.

To minimize these uncertainties and employ alternative age-dating techniques,
three different paths have been followed in the literature.

{\em a)}-- Use of different photometric systems: near-infrared
\citep[NIR;][]{bra10}, Str\"omgren \citep{gru98}, Sloan Digital
Sky Survey \citep{dic10}.

{\em b)}-- Use of different diagnostics for age/distances: luminosity function
\citep[LF][]{zoc00,ric08}, the difference in magnitude between the knee displayed
by NIR CMDs of low-mass MS stars and the MSTO as an age indicator \citep{bon10b}.

{\em c)}-- Use of advanced evolutionary phases: asteroseismology of cluster red
giant stars \citep{mig12}, and white dwarf (WD) cooling sequences \citep{ric06,deg09,bed09}.

These alternative methods are important since they provide independent constraints
on potential systematic errors affecting the 'classic' age determinations.
Some of these techniques are themselves weakly affected by uncertainties either
in distance modulus and reddening, or in the model input physics, or in the
assumptions on still uncertain physical mechanisms
(e.g., efficiency of atomic diffusion, super--adiabatic envelope convection)
that determine the calibration of the MSTO clock.

The identification of new diagnostics becomes even more relevant if they involve
advanced evolutionary phases. Absolute cluster ages \citep{ric08} and distances
\citep{zoc01} based on WDs, in spite of the intrinsic faintness of these 'clocks',
have provided this field with a fresh new perspective. Cluster WDs bring
forward two decisive positive features:
{\em i)}-- their cooling is virtually independent of the progenitor initial
metal content \citep[see, e.g.][]{sal10} and  
{\em ii)}-- they are the endpoint of the evolution of the large majority of stars,
and their properties also provide the opportunity to constrain the modeling of
%
%
their progenitor evolution \citep[][and references therein]{alt10a}. 

In the following, we discuss new evolutionary features that we identified
along the WD isochrones in the NIR bands (Bono 2010, but see also Salaris et al.~2000).
These features are located at faint magnitudes
($M_{\rm K}\approx$13--14 mag) and their detection in GCs
($m_{\rm K}\approx$26--30 mag) is beyond the capabilities of current
NIR detectors available at the 10m class telescopes. However, this
magnitude range will become a unique playground for the next
generation of extremely large telescopes (ELTs), namely the
European-ELT [E-ELT]\footnote{http://www.eso.org/public/teles-instr/e-elt.html},
the Thirty Meter Telescope [TMT]\footnote{http://www.tmt.org/}, and the Giant
Magellan Telescope [GMT]\footnote{http://www.gmto.org/}.
In \S 2 we present the theoretical framework and discuss WD isochrones
and LFs both in the optical and in the NIR bands.
In the last section we outline the impact that the new generation of
ELTs will have on deep and precise NIR photometry of crowded fields.

\section{Theoretical framework}

Fig.~1 displays BaSTI isochrones at solar metallicity (scaled solar metal
mixture) for 10, 12, and 14~Gyr old populations, including MS, TO, red
giant branch and the DA (H-atmosphere) WD cooling sequence
\citep{pie04,sal10}. The left panel shows the isochrones in the (V,V-I) CMD.
We used here updated bolometric corrections for the WD models,
employing the recent results by \citet{tre11}  that include the
Lyman-alpha profile calculations from \citet{kow06}.
The WD isochrones display a clearly defined blue TO (WDBTO, diamonds)
whose magnitude and color are correlated with cluster age.
The onset of the WDBTO is caused by the pile-up
at the bottom of the WD sequence of increasingly more massive
-- hence with smaller radii -- structures. These WDs originate
from higher mass progenitors that moved onto the WD sequence during
earlier stages of the cluster evolution.

If the distance is independently known (i.e., by means of horizontal branch
fitting, MS fitting, RR Lyrae variables), a match of the WDBTO
\citep[determined from the magnitude of the LF cutoff in either
the I- or the V band; see, e.g.,][]{han04,bed09} with models provides
the cluster age.  If the distance is not known, the difference in
magnitude between the MSTO (triangles) and the WDBTO can
also be adopted to constrain the absolute age, because
the bottom of the WD sequence becomes fainter with age faster
than the MSTO. Overall, this latter magnitude difference increases
with age by $\sim$0.1~mag/Gyr in V for ages between 10 and 14~Gyr.
In the widely employed I-band, the same parameter scales as $\sim$0.07~mag/Gyr.
The color difference between MSTO and WDBTO can in principle also be
used, but given present observational \citep[the large color spread
at the faint limit of the observed WD sequences in GGCs; see, e.g.,][]{bed09}
and theoretical uncertainties (colors and ${\rm T_{eff}}$ for stars with 
convective envelopes), it is the magnitude difference that is the
better-suited observable.

The key advantages of this 'differential' approach are the following:

{\sl a)} it is independent of uncertainties on the cluster distance and the reddening;

{\sl b)} it is minimally affected by uncertainties in the photometric zero-point calibrations;

{\sl c)} the position of the WDBTO is minimally affected by the progenitors' chemical composition;

{\sl d)} it can be applied over a broad range of cluster ages;

{\sl e)} it is minimally affected by incompleteness at the faint limiting magnitudes,
as long as the WDBTO can be unambiguously identified.
The are two main drawbacks:

{\sl a)} linear and accurate photometry is required over a range of $\approx$12 mag and 

{\sl b)} the WDBTO is located at very faint  magnitudes. This means that even by using
ACS/WFPC3 onboard the HST, this approach can only be applied to a few nearby globulars.

\begin{figure}[ht!]
\hspace*{1truecm} \includegraphics[height=0.60\textheight,width=0.43\textwidth]{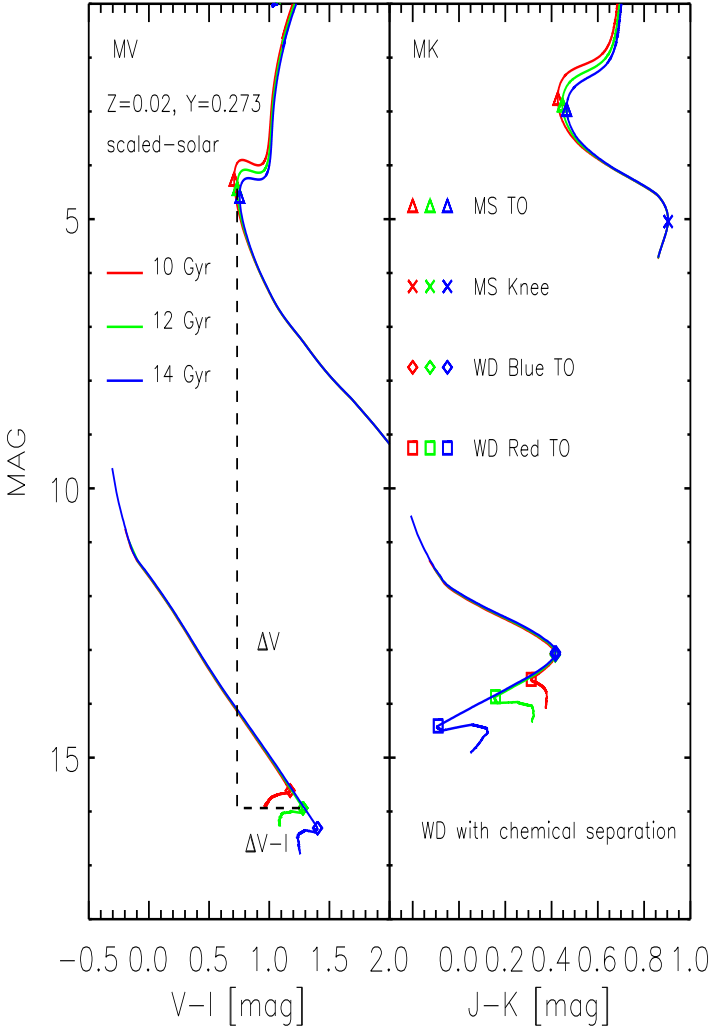}
\caption{Left -- Predicted (V,V-I) CMD for cluster isochrones at solar chemical
composition (see labeled values) and scaled solar chemical mixture. The red, green, 
and blue solid lines in the top right side of the panel show isochrones of 10, 12, 
and 14 Gyr (BaSTI data base). The triangles mark the MSTO.
The red, green, and blue solid lines in the bottom left side of the panel show WD
isochrones of 10, 12, and 14 Gyr. The WD isochrones account for chemical separation
\citep{sal10}. The diamonds mark the WDBTO.
The vertical and horizontal black dashed lines display the difference in magnitude
and in color between the 12 Gyr MSTO and the WDBTO.
Right -- Same as left, but in the NIR (K,J-K) CMD. The triangles mark the MSTO,
the crosses mark the MS knee \citep{bon10b}, the diamonds mark the WDBTO, while
squares mark the WDRTO.
}\label{optnir}
\end{figure}

The right-hand panel of Fig.~1 displays the same sets of isochrones, but in the
NIR (K, J-K) CMD.  Here the morphologies of the isochrones are more complicated 
because of the increasing  contribution by collision-induced absorption (CIA) of
molecular hydrogen \citep{han98, sau99} to the opacity in the atmospheres.
The strong infrared bands of CIA greatly reduce the infrared flux and increase
the flux at shorter wavelengths, producing a typical turn to the blue of NIR
colors with decreasing ${\rm T_{eff}}$.
As a consequence, the lower MS shows a clearly defined knee (position marked with
asterisks), that is, at fixed chemical composition, independent of the cluster
age. The difference, either in magnitude or in color, of this feature with respect to
MSTO stars can also be adopted to constrain the absolute age of star clusters
\citep{bon10b}.

The effect of the CIA opacity on the colors of WD models is similar, but the
resulting WD isochrones display an even more complex behavior
than MS isochrones. To constrain their properties we followed, a 14~Gyr 
WD isochrone in the K,J-K CMD at fixed chemical composition together with the
underlying WD cooling tracks for stellar masses ranging from M=0.55 to
1M$_\odot$, see the top panel of Fig.~2. Models plotted in this panel
show that WD cooling tracks of increasing mass turn to the blue
at almost the same color (corresponding to ${\rm T_{eff}}\approx$5000 \ K)
and increasingly fainter K-band magnitudes. The WD isochrone (red line) at 
first closely follows a single-mass WD cooling track, 
%
%
the value of this mass being fixed by the adopted initial-to-final mass 
relation (IFMR). The adopted semi-empirical IFMR is taken from Eq.~1 in 
Salaris et al. (2009) for initial masses above 0.88$M_{\odot}$, while for 
lower mass values we adopted a constant value of the final WD mass equal 
to 0.54$M_{\odot}$ (consistent with the empirical mass determination 
of bright WDs in the GC M4 by \citet{kal09}). 
The blue TO that we see in the NIR
WD isochrones is in this case caused by the onset of CIA along this single
mass cooling track. At fainter magnitudes the WD isochrone becomes populated
by progressively more massive WDs. However, these stellar structures are
increasingly less blue because of the different cooling speed and onset 
time of the CIA.
This means that after the blue TO, an old NIR WD isochrone also displays
a somewhat red TO (WDRTO) due to the appearance of the more massive objects
(Bono 2010, but see also Salaris et al.~2000).

\begin{figure}
\includegraphics[height=0.45\textheight,width=0.38\textheight]{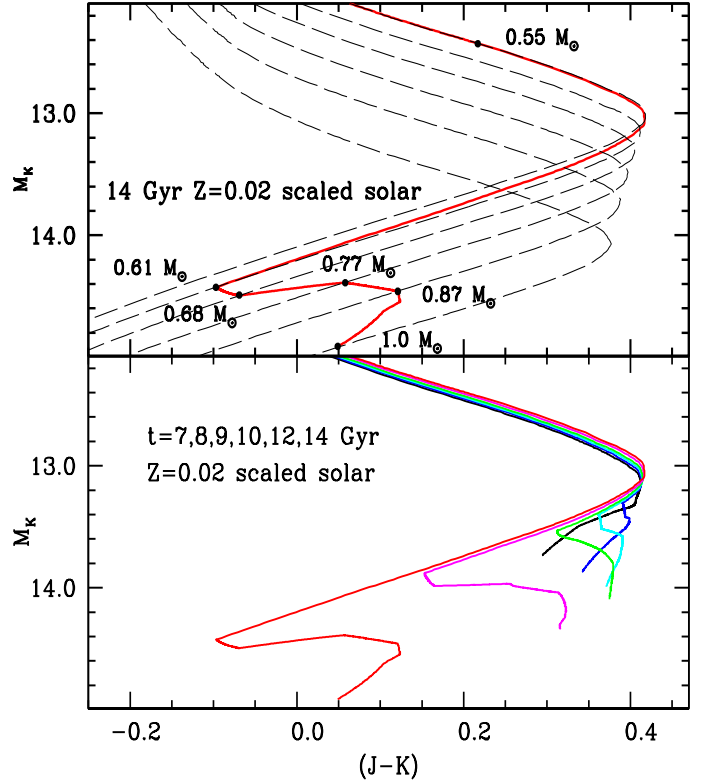}
\caption{Top -- Near-Infrared (${\rm M_K}$, J-K) -- color-magnitude diagram
for a set of white dwarf cooling tracks (long dashed lines) at solar
chemical composition (Z=0.02) and scaled solar chemical mixture.
The cooling tracks range from M=0.55  to 1.0 $M_\odot$. The red solid
line shows the WD isochrone for 14 Gyr, while the black filled circles
mark different mass values along the isochrone.
Bottom -- Same as top, but for a set of white dwarf isochrones.
Lines of different colors display isochrones ranging from 7 (black)
to 14 (red) Gyr.
}
\end{figure}

It is easy to understand why the brighter part of WD isochrones
--in any photometric system-- is always populated by essentially
single-mass WDs by recalling that at each brightness along the
cooling sequence the constraint ${\rm t_{\rm iso}=t_{\rm cool}+t_{\rm prog}}$
--where $t_{\rm iso}$ is the age of the isochrone, ${\rm t_{\rm cool}}$
the cooling time of the 'local' WD mass, and ${\rm t_{\rm prog}}$
its progenitor lifetime-- has to be satisfied. The WD cooling times
--$t_{\rm cool}$-- are very short at the bright end of the sequence,
irrespective of the WD mass, and are negligible compared to the
cluster age ($t_{\rm GC}$). Therefore, the progenitor age
--$t_{\rm prog}$-- and in turn the progenitor mass and the resulting
WD mass, has to be to a very good approximation, constant and very
close to the mass at the MSTO.
In contrast, the WD cooling time --${\rm t_{\rm cool}}$-- toward 
the faint end of the cooling sequence becomes a sizable fraction of
$t_{\rm GC}$ and the contribution of the WDs coming from higher mass
progenitors (with smaller $t_{\rm prog}$) becomes apparent in the CMD.
The exact shape of NIR isochrones is therefore determined by the
interplay between cooling times and onset of the CIA.

The bottom panel of Fig.~2 displays NIR WD isochrones for a broad age range.
Notice how for ages below $\sim$9~Gyr the WDRTO disappears. For these ages
the blue TO is caused now by the presence of more massive WDs at the bottom
of the cooling sequence, as in optical filters. In passing we note that also
in optical V, V-I CMDs one would eventually see the effect of CIA on the
isochrone colors. However, this happens at ages older than a Hubble time
for the BaSTI WD models.

\begin{figure}
\includegraphics[height=0.50\textheight,width=0.50\textwidth]{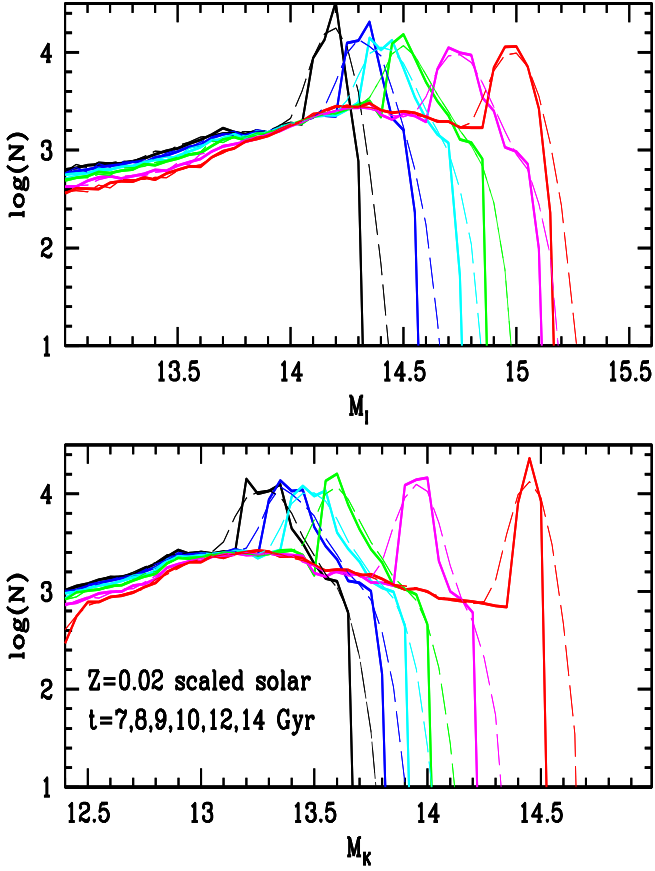}
\caption{Top -- Predicted I-band LFs at solar chemical composition
and scaled-solar chemical mixture for ages ranging from 7 to 14 Gyr (solid lines).
The faint peak at the bottom end of the LF is caused by the WDBTO.
Bottom -- Same as the panel, but for K-band LFs (solid lines). The faint peaks
are caused by the WDRTO, while the bright secondary maximum centered at
${\rm M_K}\sim$ 13.2 mag is caused by the CIA-induced blue TO (see text for more details).
The dashed lines in both panels display the same LFs, calculated from large synthetic WD
samples including a 1$\sigma$ photometric error equal to 0.05~mag in both
I and K (see text for details).}
\end{figure}

Fig.~3 shows the luminosity functions (LFs) in the I- (top) and K band (bottom)
for the same age range as displayed in Fig.~2, obtained using a Salpeter
mass function ($\alpha$=-2.35) for the WD progenitors.
The interesting feature is that the I-band LF shows a single peak connected
with the blue TO at the bottom end of the isochrones in optical filters
(see left panel of Fig.~1). 
%
%
On the other hand, the K-band LFs display a bright secondary maximum 
and a faint peak for GC ages older than 7~Gyr. The secondary maximum
($M_K\approx$13.2 mag) is the signature of the blue TO corresponding to the
onset of CIA. The position of this feature is age independent
(see also the bottom panel of Fig.~2). The faint peak is connected with 
the red TO at the bottom of the WD sequence and is caused by the
appearance of the more massive objects. 

This figure highlights several compelling features.

{\sl a)} The independence of the blue TO absolute K magnitude
--hence the location of the bright secondary maximum in the LF-- 
for old ages allows us to use this feature as a reference point. 
In particular, it can be used to determine the cluster distance, 
since it is independent of age and minimally affected by metallicity.
Operationally, observed and theoretical LFs can be shifted in magnitude
until the bright secondary maximum is matched, thus providing the apparent 
distance modulus. After this shift in magnitude is applied, the age of the 
LF can be varied until a match to the faint peak of the LF is achieved,
thus estimating the cluster age.
%
%
For a 1$\sigma$ photometric error of 0.05~mag in J/K bands  
and cluster ages older than 7~Gyr, the fits to the bright secondary 
maximum of synthetic LFs, including the quoted photometric errors, 
provide distance moduli with an accuracy better than 0.1~mag. 

The absolute K-band magnitude of the faint peak of the LF (WDRTO)
increases by $\sim$0.2~mag/Gyr for ages between 10 and 12~Gyr, and by
$\sim$0.25~mag/Gyr for ages between 12 and 14~Gyr.
It is worth noting that the absolute K magnitude of the WDRTO --i.e. the
bottom end of the LF-- is more sensitive to age than the cutoff of the LF
in either the V- or I bands.  In V the sensitivity is $\sim$0.17~mag/Gyr,
whilst in I it is $\sim$0.15~mag/Gyr in the age range between 10 and 14~Gyr.
%
%
The values of these derivatives are largely independent of the population 
initial metallicity. 

If we use the difference in K magnitude between the MSTO and the
cutoff of the WD LF as a distance-independent age indicator, the sensitivity
to age is $\sim$0.15-0.20~mag/Gyr, again higher than  the same quantity
in optical bands.

This enhanced sensitivity to age of the K-band LF can greatly improve
the internal accuracy of WD-based ages of old star clusters.
In Fig.~3 we also show the same I- and K-band LFs calculated including
a 1$\sigma$ constant photometric error equal to 0.05~mag in both filters.
This value is a conservative estimate of the accuracy achievable on GC
photometry by the Multi-Conjugate Adaptive Optics assisted instrument
MICADO. A MICADO-like camera has been selected as one of the two 
first light instruments for the E-ELT\footnote{We refer to the 
MICADO web page http://www.mpe.mpg.de/ir/micado 
and to \citet{dav10}}. Current estimates indicate a point source
sensitivity of 5$\sigma$ at limiting magnitudes of $J_{AB}\sim$31 
and $K_{AB}\sim$30 with five hours of exposure time and with 
advanced 
%
%
filters\footnote{For a more detailed discussion concerning the 
band-pass and the OH-suppression filters for MICADO@E-ELT we 
refer to \citet{gue11}.}. 
%
%
It is interesting to note that similar limiting magnitudes and a similar
sensitivity can be reached with NIRCAM@JWST with an exposure time of ten
hours\footnote{We refer to the NIRCAM web page
http://ircamera.as.arizona.edu/nircam/ and to \citet{rie11}}, but with a 
lower spatial resolution. 

%
%
By adopting the Galactic GC data base by \citet{har96} we found that this
new diagnostic could in principle be applied to more than 50\% of GCs if we 
include all globulars with apparent distance moduli smaller than 
15.0--15.5 mag. This sample includes a relevant fraction of bulge and 
inner halo GCs. 

\begin{figure}
\includegraphics[height=0.35\textheight,width=0.50\textwidth]{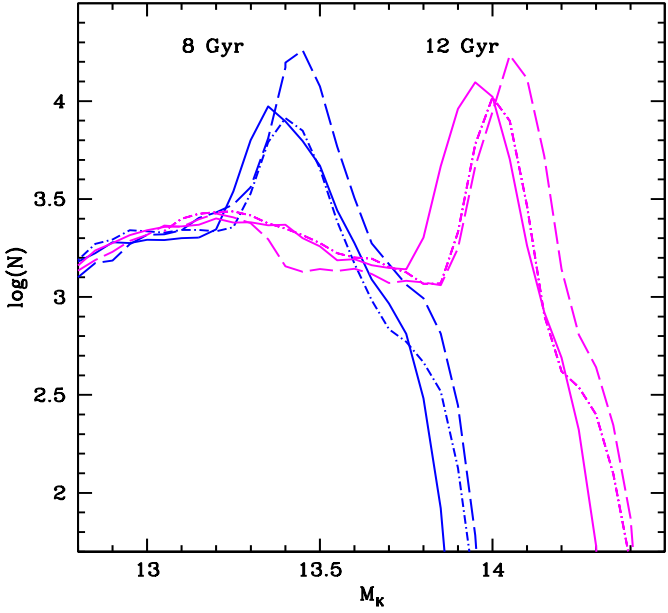}
\caption{Same as the lower panel of Fig.~3, but for two ages --8 and 12 Gyr-- 
and three different  chemical compositions. The solid lines show the LFs at 
solar chemical composition, while the dotted--dashed and long--dashed 
lines display the $\alpha$--enhanced ([$\alpha$/Fe]=0.4) LFs for two more 
metal--poor chemical compositions, [Fe/H]=-0.7 and [Fe/H]=-2.1.}
\end{figure}

{\sl b)} The dependence on the initial chemical composition
--at fixed IFMR-- is extremely weak. The location of the red TO 
--estimated as the faint peak in the LF-- is also weakly affected 
by the metallicity. The uncertainties affecting the cluster 
metallicity measurements and the GC metallicity scale are typically 
smaller than 0.2 dex.
It is noteworthy that including the photometric error does not 
substantially worsen the age-resolving power of the K band LF for 
ages older than $\sim$10~Gyr.
%
%
The initial solar chemical composition adopted to discuss the WD LFs is 
typical of stellar populations and stellar systems of the Galactic disk. To
investigate the WD LFs of typical halo GCs and field stars we also adopted
two more metal--poor chemical compositions ([Fe/H]=$-$0.7 and [Fe/H]]=$-$2.1)
and an $\alpha$--enhanced metal mixture ([$\alpha$/Fe]=0.4). The WD LFs
plotted in Fig.~4 were computed by accounting for the individual progenitor
lifetimes  and demonstrate that the bright secondary maximum located at
M$_K$$\sim$13.2, caused by the CIA opacity, is independent of metallicity.
On the other hand, the faint peak in the adopted age range becomes 0.05 mag
fainter when moving across the metallicity range of halo GCs
(-0.7 $\lesssim$ [Fe/H]]$\lesssim$ -2.1). The difference in luminosity attains 
a very similar value when moving from metal--rich GCs ([Fe/H]=-0.7) to a solar 
chemical composition. The decrease in luminosity of the faint peak arises 
because a decrease in metallicity causes a decrease in the evolutionary 
lifetime of the WD progenitors. This means that the more metal-poor WD sequences 
at fixed cluster age and IFMR reach cooler effective temperatures and in 
turn fainter magnitudes than more metal--rich sequences. This evidence further 
supports the minimal dependence on chemical composition of WD LFs, and indeed 
a difference of more than two dex in iron abundance causes a difference in 
the K-band LF of at most 0.1 mag. This difference corresponds for old 
stellar populations to much less than 1~Gyr in terms of age estimates.

{\sl c)} The position in the NIR CMD of the blue TO depends only on the IFMR
\citep[see, e.g.][and references therein]{kal09,sal09}. Changes in the
evolving WD mass in this region from 0.55$M_{\odot}$ --the value for the
IFMR adopted in BaSTI-- to 0.61$M_{\odot}$ (a considerable change, and 
probably an overestimate of the final mass for typical MSTO masses in GCs)
increases the K magnitude of the blue TO by $\sim$0.1~mag.
If extremely accurate distances are available independently of
the cooling sequence, the observed absolute magnitude of this feature
may in principle be used to constrain the IFMR for low-mass metal-poor stars.

{\sl d)} The reason for the enhanced sensitivity to age of the K-band WD LF is
disclosed by Fig.~5, which shows the run of the bolometric corrections to
the I- and the K-band (${\rm BC_I}$ and ${\rm BC_K}$) along the fainter portion
of a representative cooling track of a 0.55$M_{\odot}$ model. In general, 
the ${\rm BC_K}$ values have a much steeper dependence on ${\rm T_{eff}}$
(hence bolometric luminosity) than ${\rm BC_I}$; the local maximum of
${\rm BC_K}$ at ${\rm T_{eff}}\sim$5000~K marks the onset of CIA opacity.
Beyond  this maximum, ${\rm BC_K}$ drops very fast with decreasing
${\rm T_{eff}}$, whilst ${\rm BC_I}$ continues to steadily increase
with decreasing ${\rm T_{eff}}$, with a shallower slope.
Taking into account this behavior of the bolometric corrections, and
given that ${\rm M_x}$ in a generic photometric band X can be written as
${\rm M_x=M_{bol}-BC_X}$, the dimming rate of ${\rm M_I}$ will
necessarily be slower than for ${\rm M_K}$, because of the opposite
effect of the corresponding bolometric corrections on the decrease
in the bolometric luminosity with time.

%
%
Another important consequence of this enhanced age-sensitivity is that the
effect of CO separation upon crystallization will induce a larger difference 
--with respect to the case of no separation-- in the magnitude of the bottom 
end of the K-band LF (for a given WD isochrone age) compared to the case of 
filters at optical wavelengths. This is because CO separation causes a 
slower WD cooling, hence brighter WDs at a fixed (old) population age.
Because of the behavior of the bolometric corrections discussed above  
(see Fig.~5), a given difference in bolometric luminosity of the termination 
of the WD sequences translates into a larger magnitude difference for the 
K-band compared to optical bandpasses.

\begin{figure}
\includegraphics[width=0.50\textwidth]{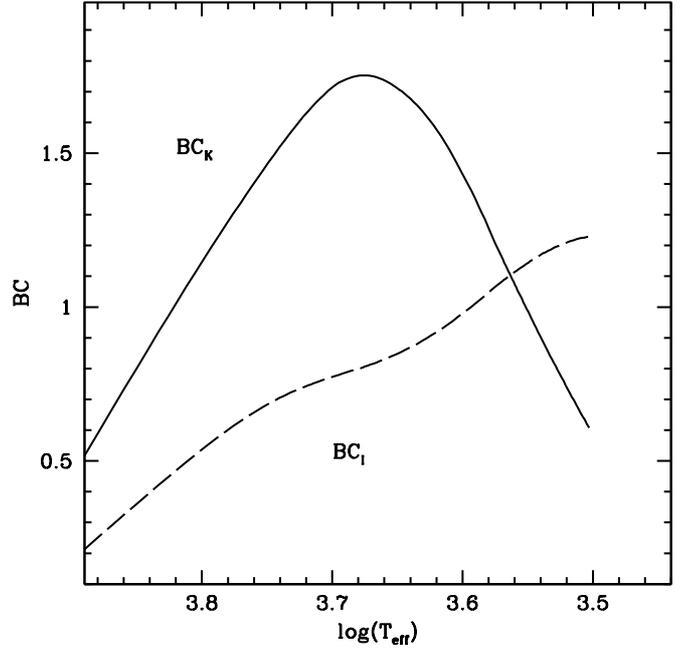} \label{fig:BC}
\caption{Bolometric corrections to the K- and I band as a function of ${\rm T_{eff}}$
along the 0.55$M_{\odot}$ cooling track. The maximum of the ${\rm BC_K}$ located at
${\rm T_{eff}}\sim$ 5000~K marks the onset of CIA opacity.}
\end{figure}

\section{Conclusions and final remarks}

The use of the next generation of ground-based ELTs and of new space
facilities (GAIA, JWST) will bring forward crucial advances in our
understanding of the Universe.
Whenever either a new instrument, a new wavelength region, or a new observing
facility becomes available we are at a classical crossroads, with two alternatives
to follow:
{\em i)}-- improve the precision of well-known diagnostics (semantic approach);
{\em ii)}-- develop new diagnostics (archetypal approach). The latter route is
less trivial, but offers a major reward: the opportunity of identifying new
evolutionary features that can provide independent checks on the systematics
affecting well-established distance and age indicators.

This latter path is what was followed in the analysis of the inflection point
along the low-mass MS in NIR CMDs \citep{bon10b}. This point is marginally
dependent on metallicity and independent of reddening, and can be adopted 
to provide robust estimates of both cluster ages and distances. The same 
applies to both blue and red TOs predicted in NIR WD isochrones and LFs, 
as discussed in this investigation.  Once again, they can be adopted
to constrain both the cluster ages and distances. Using different
diagnostics to constrain the properties of the same stellar system has
several decisive advantages:

{\em i)} --Constraints on systematics--
Using additional evolutionary phases provides the opportunity of 
constraining the input physics and the physical assumptions adopted to
construct models for the MS and advanced evolutionary stages.
Moreover, diagnostics relying on different evolutionary phases and based
on different observables (CMDs, LFs) can be used to constrain the
precision of photometric zero-points and model atmosphere adopted to
transform the output of theoretical stellar models into the observational
plane. The GGC distances can be estimated using
several standard candles (MS fitting, tip of the RGB, RR Lyrae,
WD cooling sequences, proper motion, eclipsing binaries).
Their application to the same system can provide firm constraints
on systematics \citep{bon08,tho10}. 

{\em ii)} --Physics laboratories--
GGCs are redundant systems \citep{bon10a}, therefore their intrinsic
parameters (age, distance, chemical composition) can be estimated using
different observables. The GGC ages can be estimated using different
observables (LFs, MSTO, MS knee, WD cooling sequences). Once independent
approaches show a global concordance, the advanced evolutionary phases
can be adopted to constrain interesting physical phenomena. 
%
%
The opportunity of estimating both the cluster age and the cluster distance
using the WD cooling sequences can provide firm constraints on other age
indicators. Cluster ages based on the MSTO are still hampered by uncertainties
that affect the treatment of gravitational settling of heavy elements
\citep{ric02}. Unfortunately, recent spectroscopic measurements did not
settle this open problem \citep{the01,gra01,kor06}.
Accurate and independent age estimates will allow us to perform an empirical
calibration of the free parameters currently adopted in dealing with
atomic diffusion \citep{cha95,was10,alt10a}. 

This is the first paper of a series. The next step is to constrain the
impact that $^{22}$Ne sedimentation in high-metallicity progenitors
\citep{alt10b} and H$_2$--He CIA \citep{jor00} have on the cooling sequences of
WDs with mixed H/He atmospheric compositions  \citep{koe02,han03}.

The construction plan of ELTs is already in a mature phase, with the
E-ELT having just been approved for construction \citep{gilmo}. 
The other two projects
(TMT, GMT) are in a similar construction phase. However, we still
need to wait approximately ten years for any of them to
achieve first light. While the full application of the diagnostics
described in this paper will have probably to wait that long, clearly 
the current and next generation of instruments in the
10m class telescopes will play a crucial role in paving the road
for these unique facilities.

\begin{acknowledgements}
It is a real pleasure to thank L. Althaus for many useful discussions
concerning evolutionary properties of white dwarfs and R. Davies for 
several thorough discussions concerning MICADO. 
It is also a pleasure to thank the anonymous referee for his/her 
pertinent suggestions that helped us to improve the content and the 
readability of the paper. 
One of us (G.B.) thanks ESO for support as a science visitor.
This work was partially supported by PRIN-INAF 2011 "Tracing the
formation and evolution of the Galactic halo with VST" (P.I.: M. Marconi)
and by PRIN/MIUR (2010LY5N2T) "Chemical and dynamical evolution of
the Milky Way and Local Group galaxies" (P.I.: F. Matteucci).
\end{acknowledgements}


{}

\end{document}